\documentclass[twocolumn,showpacs]{revtex4}
\usepackage{times,xspace}
\usepackage{amsbsy,amssymb,amsmath,bm}
\usepackage{graphicx,color,epsfig,rotate}
\usepackage{fancyheadings}

\def\bbbc{{\mathchoice {\setbox0=\hbox{$\displaystyle\rm C$}\hbox{\hbox
to0pt{\kern0.4\wd0\vrule height0.9\ht0\hss}\box0}}
{\setbox0=\hbox{$\textstyle\rm C$}\hbox{\hbox
to0pt{\kern0.4\wd0\vrule height0.9\ht0\hss}\box0}}
{\setbox0=\hbox{$\scriptstyle\rm C$}\hbox{\hbox
to0pt{\kern0.4\wd0\vrule height0.9\ht0\hss}\box0}}
{\setbox0=\hbox{$\scriptscriptstyle\rm C$}\hbox{\hbox
to0pt{\kern0.4\wd0\vrule height0.9\ht0\hss}\box0}}}}

\newcommand{\ignore}[1]{}
\newcommand{\mComment}[1]{}
\newcommand{\gComment}[1]{}
\newcommand{\jComment}[1]{}
\newcommand{\rComment}[1]{}
\newcommand{\lComment}[1]{}

\renewcommand{\mComment}[1]{\textcolor{blue}{Manny: #1}}
\renewcommand{\gComment}[1]{\textcolor{red}{Gerardo: #1}}
\renewcommand{\jComment}[1]{\textcolor{green}{Jim: #1}}
\renewcommand{\rComment}[1]{\textcolor{magenta}{Ray: #1}}
\renewcommand{\lComment}[1]{\textcolor{purple}{Rolando: #1}}

\begin{document}

\title{The dimerized phase of ionic Hubbard models}
\author{A.A. Aligia$^{1}$ and C.D. Batista$^{2}$}
\address{Centro At\'{o}mico Bariloche and Instituto Balseiro, Comisi\'{o}n Nacional\\
de Energ\'{\i }a At\'{o}mica, 8400 Bariloche, Argentina.\\
$^{2}$Theoretical Division, Los Alamos
National Laboratory, Los Alamos, NM 87545, USA}

\date{Received \today }

\begin{abstract}
We derive an effective Hamiltonian for the ionic Hubbard model at half
filling, extended to include nearest-neighbor repulsion. Using a
spin-particle transformation, the effective model is mapped onto simple
spin-1 models in two particular cases. Using another spin-particle
transformation, a slightly modified model is mapped into an SU(3)
antiferromagnetic Heisenberg model whose exact ground state is known to be
spontaneously dimerized. 
From the effective models several properties of the dimerized
phase are discussed, like ferroelectricity and fractional charge
excitations. Using bosonization and recent developments in the theory of
macroscopic polarization, we show that the polarization is proportional to
the charge of the elementary excitations.
\end{abstract}


\maketitle





\section{Introduction}

The ionic Hubbard model (IHM) has been proposed \cite{hub,nag} for the
description of the neutral-ionic transition in mixed-stack charge-transfer
organic crystals tetrathiafulvalene-$p$-chloranil.\cite{Lemee,Horiuchi} In
the 90's the interest on the model increased due to its potential
application to ferroelectric perovskites.\cite
{Egami,res,ort,res2,gid,fab,fab2,tor} This model is a Hubbard Hamiltonian with
nearest-neighbor hopping $t$ and on-site repulsion $U$, on a bipartite
lattice with   alternating diagonal energy $\pm {\frac{1}{2}}\Delta $ for
the two sublattices. At half filling and in the atomic limit ($t\rightarrow 0
$), the ground state is a band insulator (BI) for $U<\Delta $ (the sites
with diagonal energy $-{\frac{1}{2}}\Delta $ are doubly occupied), but is a
Mott insulator (MI) for $U>\Delta $ (all sites singly occupied). The
situation is similar to the extended Hubbard model (EHM) with $\Delta =0$ but
nearest-neighbor repulsion $V$. \cite{tor,nak,tsu} In the EHM for $t=0$, as $U
$ increases, there is a transition from a charge density wave (CDW)
insulator, with alternating doubly occupied and empty sites to a MI for $U=2V
$. However, while in this model, the result persists for finite small $t$ as
it has been shown in second order perturbation theory in $t$,\cite{hirsch}
perturbation theory in the IHM becomes invalid at $U=\Delta $ and
non-trivial charge fluctuations persist even in the strong coupling limit.
Furthermore while previous results in the IHM were interpreted in terms of
only one transition, Fabrizio {\it et al.} used arguments based on field
theory, which assumes weak coupling (essentially $\Delta \ll U\ll t$),
to propose an alternative scenario that includes two transitions as $U$
is increased: the first one is an Ising-like charge transition at $U=U_{c}$
from the BI to a bond-ordered insulator (BOI). When $U$ is further increased,
the spin gap vanishes at $U_{s}>U_{c}$ due to a 
Kosterlitz-Thouless transition  between the BOI and 
the MI.\cite{fab,fab2} An intermediate BOI
phase also takes place in the EHM\ for large enough $t$. \cite{nak,tsu}

After the proposal of Fabrizio {\it et al.}, several studies of the 1D IHM
tried to identify the number and the nature of
the different phases. Torio {\it et al.} determined the phase diagram using the method of crossing of energy levels, which for this model turns out to be
equivalent to the method of jumps of Berry phases.\cite{tor}
The basic idea is that while in finite systems conventional order parameters
such as charge and spin structure factors vary continuously at a phase
transition, in a system of length $L$ one may define charge and spin
topological numbers which change discontinuously at $U_{c}(L)$ and $U_{s}(L)$.
The charge (spin) Berry phase has a step at $U_{c}(L)$ ($U_{s}(L)$), and
extrapolating these numbers to the thermodynamic limits leads usually to
accurate results for the transition lines.\cite{bos,topo} $U_{c}(L)$ might
be detected in transport experiments through some molecules or nanodevices.\cite{dots} 
Torio {\it et al.} found a phase diagram consistent with the
proposal of Fabrizio {\it et al. }for weak coupling{\it . }They also found
that for strong coupling the intermediate BOI phase persists with a width $
U_{s}-$ $U_{c}\cong 0.6t$. Recently, similar methods were used to check the
characteristics of each quantum phase transition and confirm previous results in
systems up to 18 sites.\cite{otsu} However, Monte Carlo calculations for the
bond-order parameter, polarization and localization length were interpreted
in terms of only one transition at $U_{c}$ and two phases (BI and BOI).\cite
{wil} In addition, some density-matrix renormalization-group (DMRG)
calculations have not detected the transition at $U_{s}$.\cite{brune}
Instead, other DMRG calculations using careful finite-size scaling are
consistent with three phases and two transitions and with the phase diagram
of Torio {\it et al.}, although they suggest a smaller width of the BOI
phase for small $t$ ($U_{s}-$ $U_{c}\cong 0.4t$).\cite{zha,man}

The numerical difficulties for detecting the Kosterlitz-Thouless transition
are originated in the exponential closing of the spin gap  
as $U$ approaches to $U_{s}$  from below. Consequently, direct numerical calculations 
of the relevant correlation functions are unable to detect a sharp transition at $
U_{s}$ unless a very careful finite-size scaling is made. The same difficulty
appears in the Hubbard chain with correlated hopping, where the existence
and the position of the transition is confirmed by field theory \cite{bos,jap}
and exact \cite{afq} results. In the scenario with only one transition there
is still a question about the nature of the large $U$ phase ($U>U_{c}$). Wilkens and
Martin \cite{wil} suggested that the MI phase does not exist for finite $t$, i.e.,
the spin gap remains open for any finite $U$. This is in contradiction
with the strong coupling expansion for $t\ll U-\Delta $, which maps the IHM
onto an effective spin Hamiltonian $\tilde{H}$ with closed spin gap.\cite
{nag,brune,ali} The absence of a spin gap is also supported by the {\em 
power-law decay} of charge-charge correlation functions (in spite of the
presence of a charge gap) that have been calculated numerically \cite{man} and 
analytically (using $\tilde{H}$),\cite{ali} with very good agreement between
both results. This is a consequence of the strong charge-spin coupling in the
low-energy effective theory. These results suggest that if there is only one
transition, the BOI phase should disappear from the phase diagram. We
mention here that in presence of electron-phonon interaction, the width of
the BOI phase increases and, in the adiabatic approximation, it replaces the
MI phase.\cite{fab,wil}

From the above discussion, it is clear that the existence and extension of
the BOI phase in the electronic model deserves further study, particularly
in the strong coupling limit, $t\ll U,\Delta $, since the arguments of
Fabrizio {\it et al.} are not directly applicable. In addition, little
attention has been given so far to the electric polarization or
ferroelectric properties of the BOI phase.\cite{wil,dimer} The BOI phase is
an electronically induced Peierls instability that generates a ferroelectric
state. Experimentally, a bond ordered ferroelectric state was observed in
the pressure-temperature phase diagram of the prototype compound,
tetrathiafulvalene-$p$-chloranil \cite{Lemee,Horiuchi}. In addition, as it
was pointed out by Egami {\it et al.}, \cite{Egami} the microscopic origin
of the displacive-type ferroelectric transition in covalent perovskite
oxides like BaTiO$_{3}$ is still unclear. It has been known for a long time
that a picture based on static Coulomb interactions and the simple shell
model is inadequate to describe some ferroelectric properties.\cite{Hippel}
In a recent paper,\cite{dimer} using a spin-particle transformation,\cite
{bat2,bat3} we mapped a Hamiltonian which is close to the one obtained in the strong
coupling limit of the IHM (in a sense that will become clear in Section IV)
into an anisotropic SU(3) antiferromagnetic Heisenberg model that becomes
isotropic and exactly solvable in 1D for certain combinations of the parameters. The exact
solution is a spontaneously dimerized BOI that helps to understand the physics
of the IHM in the strong coupling limit. The large value of the correlation
length $\xi =21.0728505...$ allows to understand the numerical difficulties
for detecting this phase in finite size systems. We also used a second
spin-particle transformation \cite{bat1} that maps the constrained fermions
into $S=1$ SU(2) spins and brings the effective model into a biquadratic
Heisenberg model with a spatial anisotropy along the $z$-axis which is
proportional to $\Delta -U$. This allows to interpret both transitions
and the fractional charge excitations of the BOI phase in the simple language of 
$S=1$ spins.

In this work, we derive an effective Hamiltonian $H_{eff}$ for the strong
coupling limit of the extended IHM (including $\Delta $ and $V$). As in the 
$t-J$ model, the single-site Hilbert space of $H_{eff}$ has 
three states. In two particular relevant cases
$H_{eff}$ is mapped into simple models for $S=1$ SU(2) spins. This opens the
possibility of studying both quantum phase transitions in 1D with the
quantum loop Monte Carlo technique which is specially designed to deal with
criticality due to the non-local nature of its dynamics.\cite{Jim} 
Harada and Kawashima \cite{Kawashima} showed that this technique
provides an accurate determination of the critical parameters of a
Kosterlitz-Thouless transition. The mapping to the exact solution \cite
{dimer} is briefly reviewed. We discuss the polarization of the BOI phase.
In particular, we construct the bosonized expressions for the ``twist'' operators $
z_{L}^{c}$ and $z_{L}^{s}$ \cite{res2,znos,local} which gives information on
the conducting and polarization properties of the system, and are also
related with the Berry phases. This allows us to relate the polarization in
the BOI phase with the charge of its elementary excitations.

\section{The effective Hamiltonian}

In this Section, we derive an effective Hamiltonian for the extended IHM
(EIHM) valid for $t\ll \Delta $, but any value of $U$ if $V$ is small.\cite
{note} For simplicity we restrict our study to 1D. The extension of this
derivation to dimension higher than one is straightforward but somewhat
cumbersome for $V\neq 0$. The Hamiltonian of the EIHM can be written in the
form 
\begin{eqnarray}
H &=&-t\sum_{i,\sigma }(f_{i+1\sigma }^{\dagger }f_{i\sigma }^{\;}+{\rm H.c.})+
\frac{\Delta }{2}\sum_{i}(-1)^{i}n_{i}  \nonumber \\
&+&U\sum_{i}(n_{i\uparrow }-\frac{1}{2})(n_{i\downarrow }-\frac{1}{2}) 
\nonumber \\
&&+V\sum_{i}(n_{i}-1)(n_{i+1}-1),  \label{Hamil}
\end{eqnarray}
where $n_{i\sigma }=f_{i\sigma }^{\dagger }f_{i\sigma }^{\;}$, $
n_{i}=\sum_{\sigma }n_{i\sigma }$, and $\Delta ,t>0$. This Hamiltonian has
four states per site: empty, doubly occupied, or singly occupied with spin
up or down. At half filling and for $t=0$, all odd sites (those with energy $
-\Delta /2$) have at least one particle, and no even site is doubly
occupied. Thus, the relevant low-energy Hilbert subspace ${\cal H}_{0}$ has
three states per site. Our aim is to eliminate linear terms in $t$ which mix
states of ${\cal H}_{0}$ with the rest of the Hilbert space. This can be done with 
a canonical transformation that leads to an effective Hamiltonian $H_{eff}$ which acts on $
{\cal H}_{0}$ and  contains terms up to second order in $t$. The procedure is 
completely standard (see for instance  Ref. \cite{ali}). To simplify the
construction of $H_{eff}$ , it is convenient to perform an electron-hole
transformation for the odd sites 
\begin{eqnarray}
c_{i\uparrow }^{\dagger } &=&-f_{i\downarrow }^{\;},\;\;\;c_{i\downarrow
}^{\dagger }=f_{i\uparrow }^{\;},\;\;\text{for odd }i  \nonumber \\
c_{i\sigma }^{\dagger } &=&f_{i\sigma }^{\dagger },\;\;\;\;\;\;\;\text{for
even }i.  \label{eh}
\end{eqnarray}
The Hamiltonian now becomes invariant under translation of one site ($
i\rightarrow i\pm 1$) but it does not conserve the number of particles due to the
particle-hole transformation. The $U(1)$ symmetry associated with the conservation
of the total number of electrons is now generated by a the staggered charge operator
$Q=\sum_{i,\sigma}(-1)^{i} c_{i \sigma }^{\dagger} c_{i \sigma }^{\;}$. The transformed
Hamiltonian becomes

\begin{eqnarray}
H_{2} &=&-t\sum_{i,\sigma }(c_{i+1\uparrow }^{\dagger }c_{i\downarrow
}^{\dagger \;}-c_{i+1\downarrow }^{\dagger \;}c_{i\uparrow }^{\dagger }+%
\text{H.c.})+\frac{\Delta -U}{2}\sum_{i}n_{i}  \nonumber \\
&&-V\sum_{i}(n_{i}-1)(n_{i+1}-1)+U\sum_{i}n_{i\uparrow }n_{i\downarrow }.
\label{h2}
\end{eqnarray}

Now, for $t \ll \Delta $ but arbitrary $\Delta-U$,\cite{note} we can
eliminate the states with double occupancy on any site. This constraint can
be incorporated to the fermionic algebra by defining the constrained
fermion operators: 
\begin{equation}
{\bar{c}}_{i\sigma }^{\dagger }=c_{i\sigma }^{\dagger }(1-c_{i{\bar{\sigma}}%
}^{\dagger }c_{i{\bar{\sigma}}}^{\;}),\;\;\;\;\;\;{\bar{c}}_{i\sigma
}^{\;}=c_{i\sigma }^{\;}(1-c_{i{\bar{\sigma}}}^{\dagger }c_{i{\bar{\sigma}}%
}^{\;}).
\end{equation}
Performing a canonical transformation we eliminate the terms proportional
to $t$. These terms mix the low energy subspace ${\cal H}_{0}$ with the orthogonal
high-energy subspace ${\cal H}_{1}$, which consists of all the states having
at least one doubly occupied site in this representation. Up
to second order in $t$, we obtain the effective Hamiltonian:
\begin{eqnarray}
H_{eff} &=& -t\sum_{i}({\bar{c}}_{i\uparrow }^{\dagger }{\bar{c}}
_{i+1\downarrow }^{\dagger }-{\bar{c}}_{i\downarrow }^{\dagger }{\bar{c}}
_{i+1\uparrow }^{\dagger }+\text{H.c.})  \nonumber \\
&-& V\sum_{i}(1-n_{i})(1-n_{i+1}) +E\sum_{i}(n_{i}-1) \nonumber \\
&+&\sum_{i}\hat{J}_{i}({\bf s}_{i}\cdot {\bf s}_{i+1}-\frac{1}{4}
n_{i}n_{i+1})  \nonumber \\
&+&\sum_{i\delta =\pm 1}\hat{t}_{i}^{ch}{\bar{c}}_{i+\delta \sigma
}^{\dagger }{\bar{c}}_{i-\delta \sigma }(\frac{n_{i}}{2}+2{\bf s}_{i}\cdot 
{\bf s}_{i-\delta }),  \label{eff1}
\end{eqnarray}
where $E=(\Delta -U)/2$,  and ${\bf s}_{i}$ is the spin vector at site $i$,
the components of which are $s_{i}^{\alpha }=1/2\sum_{\tau ,\tau ^{\prime }}{%
\bar{c}}_{i\tau }^{\dagger }\sigma _{\tau \tau ^{\prime }}^{\alpha }{\bar{c}}%
_{i\tau ^{\prime }}^{\;}$ with $\alpha =\{x,y,z\}$ ($\sigma ^{\nu }$ are the
Pauli matrices). The exchange interaction $\hat{J}_{i}$ and correlated
hopping $\hat{t}_{i}^{ch}$ coefficients are {\em operators} if $V\neq 0$. If 
$V=0$, $\hat{J}_{i}=2t^{2}/(U+\Delta )$ and $\hat{t}_{i}^{ch}=t^{2}/\Delta $
are constants. $\hat{J}_{i}$ comes from a second order process in which in
the original Hamiltonian, the virtual intermediate state has a doubly
occupied site at energy $+\Delta /2$ and an empty site at energy $-\Delta /2$. 
In the original language, $\hat{t}_{i}^{ch}$ represent a correlated
hopping between two second nearest-neighbor sites at energy $+\Delta /2$ ($%
-\Delta /2$) involving an intermediate state with an empty (doubly occupied
site) at energy $-\Delta /2$ ($+\Delta /2$).

For general $V$, the expressions of these operators are the following:

\begin{eqnarray}
\hat{J}_{i} &=& 2t^{2}[\frac{n_{i-1}n_{i+2}}{U+\Delta -V}+\frac{
n_{i-1}+n_{i+2}-2n_{i-1}n_{i+2}}{U+ \Delta }+
\nonumber \\
&+& \frac{(1-n_{i-1})(1-n_{i+2})}{
U+\Delta +V}]  \label{ji}
\end{eqnarray}

\begin{eqnarray}
\hat{t}_{i}^{ch} &=&t^{2}[\frac{1}{\Delta }n_{i-2}n_{i+2}+\frac{1}{\Delta +V}%
(1-n_{i-2})(1-n_{i+2}) \nonumber \\
&+&(\frac{1}{\Delta }+\frac{1}{\Delta +V})(n_{i-2}+n_{i+2}-2n_{i-2}n_{i+2})/2].  
\nonumber
\end{eqnarray}

While the general expression of $H_{eff}$ is rather complicated, it takes simpler 
forms for some particular cases. For instance, if we only keep the linear terms in $t$, what
amounts to neglect $\hat{J}_{i}$ and $\hat{t}_{i}^{ch}$, the resulting $
H_{eff}^{a}$ already contains the non-trivial charge fluctuations of the
EIHM near both transitions and is the minimal model to describe the system
in the strong coupling limit. We can rewrite $H_{eff}^{a}$ as a spin 1  model
by using the spin-particle transformation \cite{bat1}

\begin{eqnarray}
S_{j}^{+} &=&\sqrt{2}\ (\bar{c}_{j\uparrow }^{\dagger }\
K_{j}+K_{j}^{\dagger }\ \bar{c}_{j\downarrow }^{\;}),  \nonumber \\
S_{j}^{-} &=&\sqrt{2}\ (K_{j}^{\dagger }\ \bar{c}_{j\uparrow }^{\;}+\bar{c}%
_{j\downarrow }^{\dagger }\ K_{j}),  \nonumber \\
S_{j}^{z}\ &=&\bar{n}_{j\uparrow }-\bar{n}_{j\downarrow },
\end{eqnarray}
where $K_{j}^{\;}$ is the kink operator \cite{bat1}: 
\begin{equation}
K_{j}^{\;}=\exp [i\pi \sum_{\substack{k<j}}\ \bar{n}_{k}]\ .
\end{equation}
One way to visualize the effects of this transformation is to note the
corresponding mapping of the fermion states at any site to the spin states $%
|SS^{z}\rangle $: $|j0\rangle \rightarrow |j10\rangle $, ${\bar{c}}%
_{j\uparrow }^{\dagger }K_{j}^{\;}|0\rangle \rightarrow |j11\rangle $, and ${%
\bar{c}}_{j\downarrow }^{\dagger }K_{j}^{\;}|0\rangle \rightarrow
|j1-1\rangle $. Up to an irrelevant constant, the resulting $H_{eff}$ to first order in $t$ is

\begin{eqnarray}
H_{eff}^{a} &=&
\sum_{i}[t(S_{i}^{x}S_{i+1}^{x}+S_{i}^{y}S_{i+1}^{y})P_{i,i+1}
\nonumber \\
&-& V (S^z_{i})^{2} (S^z_{i+1})^{2} \; +(E+2V) (S_{i}^{z})^{2}],
\label{spin1}
\end{eqnarray}
where $P_{ij}$ projects over the states with $S_{i}^{z}+S_{j}^{z}=0$ (one
may use $P_{ij}=1-(S_{i}^{z}+S_{j}^{z})^{2}$ in Eq. (\ref{spin1}) ).

When the second order terms are included for $V=0$ (as in
the original IHM), the effective spin Hamiltonian becomes:

\begin{eqnarray}
&&H_{eff}^{b}=%
\sum_{i}[t(S_{i}^{x}S_{i+1}^{x}+S_{i}^{y}S_{i+1}^{y})P_{i,i+1}+E(S_{i}^{z})^{2}
\nonumber \\
&+&\frac{J}{8}%
\{(S_{i}^{+}S_{i+1}^{-})^{2}+(S_{i+1}^{+}S_{i}^{-})^{2}+2S_{i}^{z}S_{i+1}^{z}-2(S_{i}^{z})^{2}(S_{i+1}^{z})^{2}\}
\nonumber \\
&-&\frac{t_{ch}}{4}\sum_{\delta =\pm 1}(S_{i+\delta }^{z})^{2}\;(S_{i+\delta
}^{x}S_{i-\delta }^{x}+S_{i+\delta }^{y}S_{i-\delta }^{y})(S_{i-\delta
}^{z})^{2}  \nonumber \\
&\times &\!\!\!\{(S_{i}^{+}S_{i-\delta }^{-})^{2}+(S_{i-\delta
}^{+}S_{i}^{-})^{2}  \nonumber \\
&&+2S_{i}^{z}S_{i-\delta }^{z}+2(S_{i}^{z})^{2}(S_{i-\delta }^{z})^{2}\;\}],
\label{spin1b}
\end{eqnarray}
where $J=2t^{2}/(U+\Delta )$ and $t_{ch}=t^{2}/\Delta .$

$H_{eff}^{a}$ is the minimal model that describes the physics of the EIHM in
the strong coupling limit. It can be studied with quantum Monte Carlo or
DMRG and its solution can bring further insight into the nature of the BOI
phase, the precise position of the phase transitions, and the effect of $V$
on them. Note that these spin 1 Hamiltonians as well as Eq. (\ref{s1inv})
below, conserve not only total $S^{z}$, but also the original charge 
$Q=\sum_{i}(-1)^{i} (S^{z}_{i})^{2}|$, and the ground state is in the $Q=0$
subspace. This seems to be an important reduction of the relevant Hilbert
space for numerical calculations.

\section{Limit of large $V$}

For $V\gg t$, the ground state of the EIHM\ can be obtained by
second-order perturbation theory in $t$, extending previous results for the
EHM.\cite{hirsch} One can start either from $H$ or $H_{eff}$. In the latter
case, one can easily see that $\hat{t}_{i}^{ch}$ becomes irrelevant and that
there is an additional contribution $J_{ad}=2t^{2}/(U-\Delta -V)$ to $J$
when empty sites are eliminated by another canonical transformation.

Starting from a CDW with maximum order parameter, the energy per site up to
second order in $t$ becomes

\begin{equation}
E_{II}=-E-V-\frac{t^{2}}{\Delta +3V-U},  \label{ii}
\end{equation}
where the subscript II means ``ionic insulator'' to distinguish it from the
BI because correlation effects due to $V$ are also present. In the MI phase,
one has an effective Heisenberg model with exchange interaction $J_{tot}$
given by Eq.(\ref{ji}) for all $n_{l}=1$ plus $J_{ad}$. From the exact
solution, the ground state energy of this model is $-J_{tot}\ln 2$,\cite{bah}
and therefore

\begin{equation}
E_{MI}=-J_{tot}\ln 2\text{; }J_{tot}=\frac{4\Delta t^{2}}{[(U-V)^{2}-\Delta
^{2}]}  \label{mi}
\end{equation}
For $t=0$, the transition between both phases is at $U=\Delta +2V$. At this
point, when $V\gg t$, the perturbative corrections in $t^{2}$ converge and the
charge fluctuations of both phases have an energy cost $V\gg t$. Consequently,
the essential ingredient which can eventually lead to the
formation of the BOI phase is absent. Therefore, there are only two phases
and one transition in this limit.

\section{Bosonization}

The previous sections and most of the paper are concerned with the
strong-coupling limit, in which $t$ is small compared either to $\Delta $ or
the interactions $U,V$. When the interactions are small, usually the
continuum limit field theory plus renormalization group allows a precise
description of the system.\cite{voit,gog} In the EIHM, the usual procedures
fail because $\Delta $ is a strongly relevant operator. Nevertheless the
bosonized Hamiltonian was used to infer the existence of the intermediate
BOI phase in the IHM and the character of its low-energy excitations.\cite
{fab,fab2} Here we describe the theory for the EIHM plus exchange
interaction to draw conclusions to be used later, and construct the
bosonized version of the operators $U_{L}^{c,s}=\exp [i\frac{2\pi }{L}%
\sum_{j}ja(n_{j\uparrow }\pm n_{j\downarrow })]$, which enter the quantities 
$z_{L}^{c,s}=\langle g|U_{L}^{c,s}|g\rangle $ where $|g\rangle $ is the
ground state, $L$ the length of the system and $a$ the lattice parameter
(and short distance cutoff below).

\subsection{Hamiltonian and main results}

The bosonized form of $H+J\sum_{i}{\bf s}_{i}\cdot {\bf s}_{i+1}$ is \cite
{bos,voit,schu}

\begin{equation}
H=H_{c}+H_{\sigma }+H_{c\sigma },  \label{hbos}
\end{equation}
where $c$ ($\sigma $) denotes charge (spin), $H_{\nu }$ ($\nu =c,\sigma $)
is a sine-Gordon Hamiltonian 
\begin{eqnarray}
H_{\nu } &=&\int dx\{\frac{v_{\nu }}{2}[\pi K_{\nu }\Pi _{\nu }^{2}(x)+\frac{%
(\partial _{x}\phi _{\nu })^{2}}{\pi K_{\nu }}]  \nonumber \\
&&+\frac{m_{\nu }}{a^{2}}\cos (\sqrt{8}\phi _{\nu })\},  \label{sg}
\end{eqnarray}
and the charge-spin interaction is

\begin{equation}
H_{c\sigma }=\int dx\frac{2\Delta }{\pi a}\cos (\sqrt{2}\phi _{c})\cos (%
\sqrt{2}\phi _{\sigma }).  \label{ros}
\end{equation}
In terms of the density operators $\nu _{kr}$ for particles with wave vector 
$k$ moving in the $r$ (+ or -) direction, the phase fields are \cite{schu}

\begin{eqnarray}
\phi _{\nu }(x) &=&-\frac{i\pi }{L}\sum_{k\neq 0}\frac{1}{k}e^{-\eta
|k|x-ikx}(\nu _{k+}+\nu _{k-})-\frac{N_{v}\pi x}{L},  \nonumber \\
\Pi _{\nu }(x) &=&\frac{1}{L}\sum_{k\neq 0}e^{-\eta |k|x-ikx}(\nu _{k+}-\nu
_{k-})-\frac{J_{v}}{L},  \label{fiel1}
\end{eqnarray}
where

\begin{eqnarray}
N_{c,\sigma } &=&[(N_{+\uparrow }+N_{-\uparrow })\pm (N_{+\uparrow
}+N_{-\uparrow })]/\sqrt{2},  \nonumber \\
J_{c,\sigma } &=&[(N_{+\uparrow }-N_{-\uparrow })\pm (N_{+\uparrow
}-N_{-\uparrow })]/\sqrt{2},  \label{fiel2}
\end{eqnarray}
and $N_{r\sigma }$ is the operator of the total number of particles with
spin $\sigma $.

The parameters entering Eqs. (\ref{sg}) are determined by the following
equations in terms of the usual g-ology coupling constants:

\begin{eqnarray}
v_{\nu } &=&\sqrt{(v_{F}^{\nu })^{2}-(\frac{g_{\nu }}{2\pi })^{2}}%
,\;\;\;m_{c}\;=\;\frac{g_{3\perp }}{2\pi ^{2}},\;\;\;m_{\sigma }\;=\;\frac{%
g_{1\perp }}{2\pi ^{2}},  \nonumber \\
v_{F}^{\rho }\; &=&\;2ta+\frac{g_{4||}+g_{4\perp }}{\pi },\;\;\;v_{F}^{%
\sigma }\;=\;2ta+\frac{g_{4||}-g_{4\perp }}{\pi },  \nonumber \\
g_{c }\; &=&\;g_{1||}-2g_{2||}-2g_{2\perp },\;\;\;g_{\sigma
}\;=\;g_{1||}-2g_{2||}+2g_{2\perp },  \nonumber \\
K_{\nu }\; &=&\;\sqrt{\frac{2\pi v_{F}^{\nu }+g_{\nu }}{2\pi v_{F}^{\nu
}-g_{\nu }}}.  \label{mass}
\end{eqnarray}
The masses entering Eqs. (\ref{sg}) are proportional to the Umklapp coupling 
$g_{3\perp }=a(U-2V+3J/2)$ and the backward-scattering one $g_{1\perp
}=g_{\sigma }\;=a(U-2V-J/2)$, while $g_{c}<0$ for positive $U$, $V$, and $J$.%
\cite{bos} For $\Delta =0$, the charge and spin sectors are decoupled and the
usual renormalization group procedure for the sine Gordon model can be
applied. By increasing $U$, there is a transition from the II (CDW) to
the BOI at $U_{c}=2V-3J/2$, and another one from the BOI to the MI at $%
U_{s}=2V+J/2$.\cite{bos} 
Including second-order corrections to the coupling constants, it has been shown 
that for $J=0$ the BOI phase still exists for weak $U \sim 2V$.\cite{tsu}
The sign of $g_{3\perp }$ ($g_{1\perp }$)
determines the charge (spin) sector. The width of the BOI phase goes to zero
with $J$.

For $\Delta \neq 0$, $V=J=0$, Fabrizio {\it et al.} \cite{fab} replaced $%
H_{c}$ by an effective free massive theory for the MI\ phase. Then, treating 
$H_{c\sigma }$ perturbatively, they obtained an effective $H_{\sigma }$ with
a renormalized $m_{\sigma }$, showing that lowering $U$, a spin gap opens
before the charge transition and therefore a BOI phase exists between the
two transitions. They also discussed the excitations of the BOI phase taking
the interaction part of the Hamiltonian as a phenomenological
Ginzburg-Landau energy functional $F$ with effective interactions. For its
importance in what follows we briefly review this analysis. Ignoring some
constants and factors and dropping the subscript $\perp $, we can write 
\begin{eqnarray}
F &=&{ g}_{3}\alpha _{c}^{2}+{ g}_{1}\alpha _{s}^{2}+ { \Delta} \alpha _{c}\alpha _{s},
\nonumber \\
\alpha _{v} &=&\cos (\sqrt{2}\phi _{v}).  \label{f}
\end{eqnarray}
In Eq. (\ref{f}) and in the rest of this subsection, the $g_i$ and $\Delta$ are 
effective parameters renormalized by the interactions and their numerical
values are different from those given previously.
Since we are in the spin sector of the BOI phase, this implies that the spin
gap is open and therefore the effective $g_{1}<0$. Then, $|\alpha _{s}|=1$
minimizes the energy. If also $g_{3}<0$ (corresponding to the BI phase) the
energy is minimized by either $\alpha _{c}=1$, $\alpha _{s}=-1$ or $\alpha
_{c}=-1$, $\alpha _{s}=1$. If one has a soliton (topological excitation) in
the system between these two vacua, clearly the jump in both phases is $%
\Delta \phi _{v}=\pm \pi /\sqrt{2}$. Since the smooth part of the charge $%
\rho (x)$ (spin $S_{z}(x)$) operator is $-\sqrt{2}\partial _{x}\phi _{c}/\pi 
$ ($\partial _{x}\phi _{s}/(\sqrt{2}\pi) $) this excitation has total charge 1
and spin 1/2, as expected. If $g_{3}>2\Delta $, the minima occur 
for $\alpha_{c}=\Delta /(2g_{3})$, $\alpha _{s}=-1$ or $\alpha _{c}=-\Delta /(2g_{3})$, 
$\alpha _{s}=1$. For an excitation between vacua with different $\alpha _{s}$
the spin is 1/2 as before, but now the smallest change in the charge field
is $\Delta \phi _{c}=\pm C_{1/2}(\pi /\sqrt{2})$, being $C_{1/2}=1-2\arccos
[\Delta /(2g_{3})]/\pi $ the fractional charge of the excitation. As $g_{3}$
increases, $C_{1/2}$ decreases. This is consistent with a continuous
transition to the MI phase, where the elementary excitations are spinons
without charge. In addition there is a pure charge excitation between the
two possible values of $\phi _{c}$ modulo $2\pi $ for the same values of $%
\alpha _{c}$ and $\alpha _{s}$. Its charge is $C_{0}=1-C_{1/2}$.

\subsection{The displacement operators}

The quantity $z_{L}^{c}=\langle g|U_{L}^{c}|g\rangle $ where $U_{L}^{c}=\exp
[i\frac{2\pi }{L}X]$ where $L$ is the length of the system and $%
X=\sum_{j}ja(n_{j\uparrow }+n_{j\downarrow })]$, was first proposed by Resta
and Sorella as an indicator of localization in extended systems.\cite{res2}
Using symmetry properties, Ortiz and one of us  showed that for
translationally invariant interacting systems with a rational number $n/m$
of particles per unit cell the correct definition is $\langle
g|(U_{L}^{c})^{m}|g\rangle $ \cite{znos}. This definition was used to characterize
metal-insulator and metal-superconducting transitions in one dimensional
lattice models. In the thermodynamic limit, $L\rightarrow \infty $, $|z_{L}^{c}|$ is 
equal to 1 for a metallic periodic system and to 0 for the insulating state.
This is also true for non-interacting disordered systems.\cite{local} In the
thermodynamic limit, $z_{L}^{c}$ provides a way to calculate the expectation
value of the position operator (related to the macroscopic electric
polarization by $P=e\langle X\rangle /L$) and its fluctuations for a periodic
system \cite{res2,znos,local}:

\begin{eqnarray}
\lim_{L\rightarrow \infty }\frac{e}{2\pi m}{\rm Im}\ln z_{L}^{c} &=&\frac{e}{%
2\pi m}\gamma _{c}=P-P_{0}\ ,  \nonumber \\
-\lim_{L\rightarrow \infty }\frac{L^{2}}{(2\pi m)^{2}}\ln |z_{L}^{c}|^{2}
&=&\langle X^{2}\rangle -\langle X\rangle ^{2},  \label{berry}
\end{eqnarray}
where $\gamma _{c}$ is the charge Berry phase \cite
{res,ort,tor,topo,ort2,berry} and $P_{0}$ some reference polarization. $%
\gamma _{c}$ is defined modulo $2\pi $ and therefore $P_{0}$ is defined
modulo $e/m$. These results were extended to $z_{L}^{s}$, where $X$ is
replaced by the difference between the position of up and down operators $%
\sum_{j}ja(n_{j\uparrow }-n_{j\downarrow })$, $P$ by the corresponding
difference in polarization, and $\gamma _{c}$ by the spin Berry phase $%
\gamma _{s}$.\cite{topo,berry} For calculations in finite systems, more
accurate results for $\gamma _{v}$ are obtained calculating it directly
rather than using the first Eq. (\ref{berry}).\cite{znos,topo} However, this
expression allows   a trivial calculation of $\gamma _{c}$ if all particles
in the ground state are localized.\cite{topo} In particular $\gamma _{c}=0$
for a CDW with maximum order parameter and $\gamma _{c}=\pi $ for a MI with
one particle per site. By continuity and considering that $\gamma _{c}$ can only be 0 or $\pi $ modulo $2\pi $ for a system with inversion
symmetry and one particle per site, these two values characterize the II and MI phases until the boundary of a phase transition is reached.

The difficulty in finding a bosonized expression for $U_{L}^{c,s}$ is that
the exponents are ill defined in a periodic system. Actually this is the
reason why $P$ should be defined through the Berry phase or $z_{L}^{c}$.\cite
{res2,znos,local} Nevertheless, as noted earlier,\cite{znos} $U_{L}^{c}$ is
the operator that shifts all one particle momenta by -$2\pi /L$, to the next
available wave vector to the left. Then, by inspection of the expressions of
the fields Eqs. (\ref{fiel1},\ref{fiel2}), $U_{L}^{c}$ acting on any state
increases $N_{-\sigma }$ by one and decreases $N_{+\sigma }$ also by one.\cite{note2}
Consequently, $U_{L}^{c}$ shifts $\Pi _{c}(x)$ by $-2\sqrt{2}/L$ and
leaves the other three fields invariant. From the commutation relation,
\begin{equation} 
[\phi _{c}(x),\Pi _{c}(y)]=i\delta (x-y),
\end{equation}
it follows that 
\begin{equation}
[-i\int dx\phi_{c}(x),\Pi _{c}(y)]=1
\end{equation}
Therefore the displacement operator $U_{L}^{c}$
can be constructed as an exponential of the average charge field. A similar
analysis can be carried out for $U_{L}^{s}$ and both results can be written
as \cite{note3}

\begin{equation}
U_{L}^{\nu }=\exp (i\sqrt{8}\phi _{\nu }^{a})\text{, }\phi _{\nu }^{a}=\frac{%
1}{L}\int dx\phi _{\nu }(x).  \label{uc}
\end{equation}

In the region of parameters such that the interaction $m_{\nu }$ in the sine Gordon model
Eq. (\ref{sg}) is relevant according to the renormalization group, $\phi _{\nu }(x)$ gets
frozen at the value that minimizes $m_{\nu }\cos (\sqrt{8}\phi _{\nu })$ and
the calculation of $U_{L}^{\nu }$ becomes trivial. In particular, for $
g_{3\perp }<0$ ($m_{c}<0$ corresponding to a CDW or II) we obtain $\phi
_{c}^{a}=\gamma _{c}=0$, while in the MI phase ($U\rightarrow \infty $, $m_c >0$) 
the result is 
$\sqrt{8}\phi _{c}^{a}=\gamma _{c}=\pi $, in agreement with the values 
anticipated above.\cite{topo} The difference in polarization $e/2$ modulo $e$ (see Eq. (\ref
{berry})) is consistent with the transfer of half of the electrons in one
lattice parameter which is required to convert a MI into a II in the strong
coupling limit.

In the BOI\ phase, there is a spontaneous breaking of the inversion symmetry
in the thermodynamic limit and fractional values of $\gamma _{c}/\pi $ are
allowed for $\Delta \neq 0$. The analysis of the previous subsection
indicates that in the BOI phase there are two non-equivalent values for the
frozen charge field: $\sqrt{2}\phi _{c}^{a}=\pm \arccos [\Delta
/(2g_{3})]=\pm \pi C_{0}/2$. Then, using Eqs. (\ref{berry}) and (\ref{uc})
we obtain the following relation between the polarization and the fractional
charge of the elementary spinless excitation of the BOI phase:

\begin{equation}
P-P_{0}=\pm \frac{e}{2\pi }\gamma _{c}=\pm \frac{e}{2}C_{0}.  \label{p}
\end{equation}
The sign depends on the particular ground state out of the two-fold degenerated manifold
which follows from the $Z_2$ (spatial inversion) symmetry breaking. If $P_{0}$ is 
chosen in such a way that $P=0$ for the MI (implying $P_{0}=\pm e/2$), we also obtain:

\begin{equation}
P=\pm \frac{e}{2}C_{1/2}.  \label{p2}
\end{equation}

\section{Rigorous results on an approximate effective Hamiltonian}

So far, no approximations have been made beyond that of the strong coupling
limit. In this Section we describe a simplification of $H_{eff}$ that allows
to map it into an exactly solvable model.\cite{dimer} We neglect the
correlated hopping term $t_{ch}$ and treat the exchange term $J$ as an
independent parameter. Both approximations are usually applied to 
the $t-J$ model after its derivation from the Hubbard 
model in the strong coupling limit \cite{esk,lema} in
spite of the fact that the correlated hopping term might be important for
stabilizing a superconducting  resonance-valence-bond ground state.\cite{rvb1,rvb2}

To simplify the expression for the spin Hamiltonian and unveil accidental
symmetries that appear for particular combinations of the parameters, it is
convenient to change the sign of the $s_{i}^{x}s_{i+1}^{x}$ and $%
s_{i}^{y}s_{i+1}^{y}$ term in Hamiltonian (\ref{eff1}) (this also changes
the sign of the term $(S_{i}^{+}S_{i+1}^{-})^{2}+$H.c. in 
Eq. (\ref{spin1b}) ) using the gauge transformation 
\begin{eqnarray}
&&{\bar{c}}_{i\downarrow }^{\dagger }\rightarrow -{\bar{c}}_{i\downarrow
}^{\dagger }\;\;\;\;\text{for}\;\;i=4n\;\;\text{and}\;\;i=4n+1  \nonumber \\
&&{\bar{c}}_{i\uparrow }^{\dagger }\rightarrow -{\bar{c}}_{i\uparrow
}^{\dagger }\;\;\;\;\text{for}\;\;i=4n+2\;\;\text{and}\;\;i=4n+1.
\label{gauge}
\end{eqnarray}
Then, neglecting $t_{ch}$, adding the term in $V$ in (\ref{spin1b}), and
taking $V=t$, $J=2t$, the spin 1 Hamiltonian takes the SU(2) invariant form:

\begin{equation}
H_{eff}=-t\sum_{i}[\left( {\bf S}_{i}\cdot {\bf S}_{i+1}\right)
^{2}-1]+E\sum_{i}[(S_{i}^{z})^{2}-1].  
\label{s1inv}
\end{equation}
For $E=0$, this model has an SU(3) symmetry which is hidden in this
representation. The situation is somewhat similar to the supersymmetric $t-J$
model for $J=2t$.\cite{pedro,bares} In our case, the SU(3) symmetry is
made explicit using a spin-particle transformation described in detail in
Ref. \cite{dimer} For $E=0$, the model is equivalent to an isotropic SU(3)\
antiferromagnetic Heisenberg model:

\begin{equation}
H_{eff}(E=0)=\sum_{j\in A,\mu ,\nu }t{\cal S}^{\mu \nu }(j){
\tilde{{\cal S}}}^{\nu \mu }(j+1),  \label{su3}
\end{equation}
where $A$ is an arbitrary sublattice ($j$ even or odd) in which the three
relevant states transform under the fundamental representation ${\cal S}
^{\mu \nu }$, while in the other sublattice $B$, the states transform under
its conjugate representation ${\tilde{{\cal S}}}^{\nu \mu }$. The
Hamiltonian Eq. (\ref{su3}) is invariant under staggered conjugate SU(3)
rotations, ${\cal R}$ and ${\cal R}^{\dagger }$, on sublattices $A$ and $B$
respectively and is integrable.\cite{Parkinson,Barber,Klumper} The ground
state of the exact solution is spin-dimerized and corresponds to the BOI\ in
the original language. The degree of dimerization can be characterized by
the order parameter \cite{Xian,Solyom} 
\begin{equation}
D=|h_{i-1,i}-h_{i,i+1}|,
\end{equation}
where $h_{i-1,i}=|i-1,i\rangle \langle i,i-1|$ is a projector on the SU(3)
singlet spin state, $|i-1,i\rangle =\sum_{m}|i-1m\rangle |im\rangle /\sqrt{3}
$, at the bond $(i-1,i)$. Note that $3h_{i-1,i}$ coincides with the term
with $j=i-1$ in the isotropic SU(3) Heisenberg Hamiltonian Eq. (\ref{su3}).
In the original fermion representation, $|i-1,i\rangle $ has the form

\begin{equation}
|i-1,i\rangle =\frac{1}{\sqrt{3}}(f_{k\uparrow }^{\dagger }f_{k\downarrow
}^{\dagger }+f_{i-1\uparrow }^{\dagger }f_{i\downarrow }^{\dagger
}-f_{i-1\downarrow }^{\dagger }f_{i\uparrow }^{\dagger })|0\rangle ,
\label{singlet}
\end{equation}
where $k=i-[1+(-1)^{i}]/2$ is the site of energy $-\Delta /2$ between $i$
and $i+1$. A perfect dimerized state is constructed repeating this SU(3)
singlet each two sites. There are two possibilities depending on the initial
site chosen, reflecting the Z(2) symmetry breaking of the exact solution.
The exact ground state has $D=0.4216D_{0}$ where $D_{0}$ is the value of $D$
for a perfect dimerized state.\cite{Xian} The value of the gap is rather
small ${\tilde{\Delta}=0.173178}t$, and accordingly the correlation length $%
\xi =21.0728505...$ is very large.\cite{Klumper} Therefore, one expects
that very large systems should be studied to calculate numerically the
bond-order parameter or other properties of the BOI.

The significant degree of dimerization and the form of the SU(3) singlets
Eq.(\ref{singlet}) points out an important degree of covalency in the BOI
phase even in the strong coupling limit. This is in contrast to the
solutions for the II and MI phases described previously for large $V$ (see
Section III).

While the exact solution brings important insight into the physics of a
dimerized state, one might wonder if the approximations made in the
Hamiltonian, or the parameters chosen, render the results invalid for the
IHM. If the IHM is described to first order in $t$, then $t_{ch}$ becomes
irrelevant and the question is if the BOI survives after moving $J$ and $V$
from their values at the exact solution to zero. The results presented in
Sections II and III, indicate that increasing $V$ tends to close the BOI
phase, while increasing $J$ the effect is the opposite. The effect of small
changes of $J$, $V$, and $E$ from the exactly solvable point on the energy
of the dimer state, can be calculated by first order perturbation theory. In
particular, because of the SU(3) symmetry of the ground state, the ground
state expectation value of $-({\bf s}_{i}\cdot {\bf s}_{i+1}-n_{i}n_{i+1}/4)$%
, which is a projector over an SU(2) singlet with two components, is 2/3
times the expectation value of the SU(3) singlet with three components. We
can write:

\begin{eqnarray}
\delta E_{BOI} &=&E_{BOI}-E_{BOI}^{0}=[-\frac{2}{3}(J-2t)  \nonumber \\
&&-\frac{1}{3}(V-t)]\langle h_{i,i+1}\rangle _{av}+E\langle
(S_{i}^{z})^{2}-1\rangle ,  \label{corr1}
\end{eqnarray}
where $E_{BOI}^{0}$ is the energy at the exactly solvable point and $\langle
h_{i,i+1}\rangle _{av}$ is the expectation value $\langle h_{i,i+1}\rangle $
averaged over even and odd $i$. From the energy of the exact solution, we
know that \cite{Klumper}

\begin{equation}
3\langle h_{i,i+1}\rangle _{av}=\langle \left( {\bf S}_{i}\cdot {\bf S}%
_{i+1}\right) ^{2}-1\rangle _{av}=1.796863...,  \label{aver}
\end{equation}
while because of symmetry (all spin 1 projections are equally probable) $%
\langle (S_{i}^{z})^{2}\rangle =2/3$.

Estimating the energy shifts for the other two phases as in Section II,
neglecting terms of order $t^{2}$ we have

\begin{eqnarray}
\delta E_{II} &=&E_{II}-E_{II}^{0}\simeq E-(V-t)  \nonumber \\
\delta E_{MI} &=&E_{MI}-E_{MI}^{0}\simeq -(J-2t)\ln 2\text{.}  \label{corr2}
\end{eqnarray}
Using these expressions to calculate the energies of the IHM $(V=J=0)$ for
small $E$, one obtains that $\delta E_{II}=\delta E_{MI}\cong 1.39t$ when $%
E/t=1-2\ln 2\cong -0.39$. For this value of $E$, Eq. (\ref{corr1}) gives $%
\delta E_{BOI}\cong 1.13t<\delta E_{II}$. Since we know that $E_{BOI}^{0}<$ $%
E_{II}^{0}$ and $E_{BOI}^{0}<E_{MI}^{0}$, this implies that the BOI phase is
still that of lowest energy in the strong coupling limit of the IHM if $E$
is decreased to $\sim -0.39t$. Numerical results suggest that while this
shift in $E$ is in the right direction, its magnitude is nearly two times
smaller.\cite{tor,zha,man}

\section{Polarization and fractional charge excitations}

\begin{figure}[htb]
\vspace*{0.5cm}
\includegraphics[angle=-90,width=8cm]{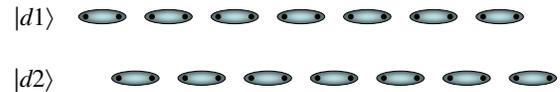}
\vspace*{-3.0cm}
\label{fig1}
\caption{Schematic plot of two dimerized ground states of the BOI.}
\end{figure}

In this section, we discuss the polarization and fractional charge
excitations of the BOI phase. To render the discussion clearer, we begin by
calculating these properties in a perfectly dimerized state with the above
mentioned SU(3) symmetry. Due to the breaking of Z(2) symmetry, there are
two such states as it is shown in Fig.\ref{fig1}. Except for a trivial normalization
constant, one of them can be written as:

\begin{equation}
|d1\rangle = \prod_{j=0}^{M-1} \Delta_j^{\dagger} |0\rangle ,  \label{d1}
\end{equation}
with $\Delta_j^{\dagger}=(f_{2j+1\uparrow }^{\dagger
}f_{2j+1\downarrow }^{\dagger }+f_{2j\uparrow }^{\dagger }f_{2j+1\downarrow
}^{\dagger }-f_{2j\downarrow }^{\dagger }f_{2j+1\uparrow }^{\dagger
})$ (the other ground state $|d2\rangle $ is obtained replacing the subscript $2j$ by $
2j+2$ in the last two terms). $M=L/(2a)$ is half the number of
sites. Using the definitions in Section IV b and applying $U_{L}^{c}$ to our perfectly dimerized ground state $|d1\rangle$ we obtain: 

\begin{equation}
\prod_{j=0}^{M-1} e^{ i\pi \frac{4j+1}{M}}
(e^{\frac{i\pi}{M}} f_{2j+1\uparrow }^{\dagger
}f_{2j+1\downarrow }^{\dagger }+f_{2j\uparrow }^{\dagger }f_{2j+1\downarrow
}^{\dagger }-f_{2j\downarrow }^{\dagger }f_{2j+1\uparrow }^{\dagger
}) |0\rangle .
\end{equation}
In the thermodynamic limit, $(\exp (i\pi /M)+2)/3\cong \exp (i \pi/3M)$, the 
mean value of $z_{L}^{c}$ is:  

\[
z_{L}^{c}(d1)=\langle d1|U_{L}^{c}|d1\rangle \cong e^{i\pi}
\prod_{j=0}^{M-1} e^{i\pi /3M} = e^{i\frac{4}{3}\pi}.
\]
This result implies a Berry phase equal to $4\pi /3$. Assuming as before that the
polarization of the MI phase is zero ($P_{0}=e/2$ modulo $e$) we have for
the polarization $P=e/6$, using Eq. (\ref{berry}) with $m=1$. This result
has a simple interpretation: the ideal dimer state is composed of isolated
``diatomic molecules''and in this case $P$ coincides with the result for one
molecule.\cite{znos} In each molecule, 1/3 of the charge (the weight of the
first term in each factor of Eq.(\ref{d1})) has been displaced from site $2j$
to $2j+1$, and since there is one molecule per two lattice parameters, the
dipolar moment has increased by $e/6$. 

It is easy to generalize this result out of the SU(3) symmetric point and
for the other perfectly dimerized state. The result is (modulo $e$)

\begin{equation}
P(d1)=-P(d2)=\frac{e}{2}\langle n_{1}-1\rangle =\frac{e}{2}\langle
1-(S_{1}^{z})^{2}\rangle \text{.}  \label{polpe}
\end{equation}
The factors that multiply $e/2$ express the weight of doubly occupied sites,
i.e., the sites with energy $-\Delta /2$. Note that as $U$ increases 
these factors decrease and they vanish when the system enters the MI phase.
In the opposite limit, at the boundary with the II phase or inside it, there
is also only one value of the polarization $P=\pm e/2$ since it is defined
modulo $e$. At this point, it is not possible to shift between two different
polarized sates by application of an electric field and the system is not
ferroelectric. Only inside the BOI phase one has ferroelectricity.

While we are unable to calculate the polarization in the ground state of the
BOI phase, our previous result Eq. (\ref{p2}) indicates that the
result for the perfectly dimerized states Eq. (\ref{polpe}) holds in the
general case. The charge of the elementary excitations of the BOI phase can
be easily determined in the spin language.\cite{dimer} Fig. 2 shows 
a schematic representation of a spinon excitation in which the two possible ground states of 
the BOI phase are separated by a spin 1 at a given site $i$. This is not an eigenstate 
because the quantum fluctuations of the spin and the bond ordered regions are
not included. However, the inclusion of these fluctuations does not change 
the charge of the soliton which is connecting both ground states. Since the original charge
operator can be written as $Q=\sum_{j}(-1)^{j}|S_{j}^{z}|$, if $S_{i}^{z}=0
$ (representing either an empty site or a doubly occupied one), the total
change in spin is zero, and the charge of the excitation is $C_{0}=\langle
(S_{l}^{z})^{2}\rangle $, where $l$ is any site. If instead $S_{i}^{z}=1$,
the spin of the excitation is 1/2 and its charge is $C_{1/2}=\langle
1-(S_{l}^{z})^{2}\rangle $. These results can be related with those of
Section IV b and are consistent with Eqs. (\ref{p2}) and (\ref{polpe}).

\begin{figure}[htb]
\vspace*{-0.5cm}
\includegraphics[angle=-90,width=8cm]{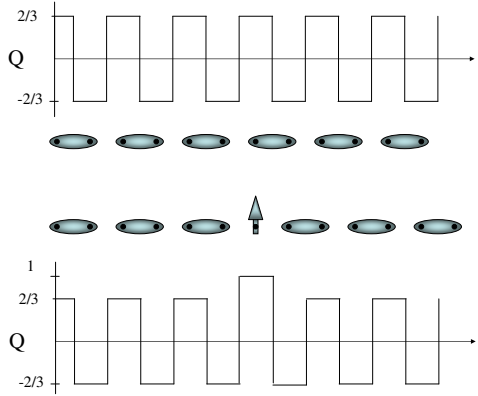}
\vspace*{0.0cm}
\label{fig2}
\caption{Schematic plot of the soliton excitation of the dimerized ground state
in the spin one representation (see Eq. (\ref{s1inv}) ). In the original fermionic 
representation, this excitation has spin $S=1/2$ and charge 
$C_{1/2}=1/3$ at the SU(3) symmetric point.}
\end{figure} 

\section{Discussion}

In summary, we have derived an effective low energy Hamiltonian for the strong 
coupling regime of an extended IHM that includes a nearest-neighbor repulsion. 
Using the spin-particle transformations introduced in Ref. \cite{bat1}, we mapped
the effective Hamiltonian into a spin one model. The spin language is useful 
to unveil hidden symmetries of the effective model for particular combinations
of the different parameters. For instance,  we showed that when the correlated
hopping $t_{ch}$ is neglected and the exchange term $J$ is treated as an
independent parameter, the effective Hamiltonian becomes a biquadratic Heisenberg
model with a single-ion anisotropy for $V=t$ and $J=2t$. In particular, the amplitude of 
the single-ion anisotropy vanishes in the region dominated by the charge
transfer instability: $U=\Delta$. At this point, the Hamiltonian can be rewritten as an 
SU(3) antiferromagnetic Heisenberg model using the general spin-particle transformations 
introduced in \cite{bat2,bat3}. The exact ground state of this 
model is a dimerized solution that corresponds to the BOI in the original fermionic
language. In the spin one language, the low energy excitations of the dimerized solution 
are spinons. The $S_z=0$ spinons carry zero spin and charge $C_0=\pm 2/3$ 
in terms of the original fermions. The $S_z=\pm 1$ spinons have spin $s=1/2$ and charge 
$C_{1/2}=\pm 1/3$ in the original fermionic version. 
Making $U \neq \Delta$ changes the value of $C_0$ and $C_1$ continuously as a function 
of $U-\Delta$ so the charge $Q$ the solitons can also take irrational values.
     
     The above described results, which were obtained in the strong coupling limit, 
are in qualitative agreement with the weak coupling bosonization approach of Fabrizio 
{\it et al}.\cite{fab2} The irrational values of the charge carried by each soliton are  
just a consequence of the asymmetry between the two sublattices introduced by
the $\Delta$ term.\cite{Rice} Since these excitations have a topological nature, it 
is expected that their characteristics will not depend on the interaction regime. 

In addition, using the bosonization procedure we showed that the 
polarization of the ferroelectric BOI is proportional to the charge 
$C_{1/2}$ of the 
elementary excitations (solitons). Therefore, the magnitude of the fractional 
charge carried by each soliton can be obtained experimentally by measuring the 
electric polarization of the ground state.

\section*{Acknowledgments}

We thank A.A. Nersesyan and D. Baeriswyl for useful discussions. This work was
sponsored by PICT 03-12742 of ANPCyT and US DOE under contract
W-7405-ENG-36. A.A.A. is partially supported by CONICET.


\begin{references}
\bibitem{note2}  We are assuming that the bottom of the band is always
filled, as usual in bosonization.
\end{references}
\end{document}